\documentclass[fleqn,usenatbib]{mnras}
\usepackage{mathptmx}
\usepackage[dvipsnames]{xcolor}
\usepackage[T1]{fontenc}

\DeclareRobustCommand{\VAN}[3]{#2}
\let\VANthebibliography\thebibliography
\def\thebibliography{\DeclareRobustCommand{\VAN}[3]{##3}\VANthebibliography}

\usepackage{graphicx}	
\usepackage{amsmath}	

\title[Testing the Galactic Centre potential with S-stars]{Testing the Galactic Centre potential with S-stars}

\author[A.~ F.~ Zakharov]{
Alexander F. Zakharov, $^{1,2}$\thanks{E-mail: {alex.fed.zakharov@gmail.com}}
\\
$^{1}$ Bogoliubov Laboratory for Theoretical Physics, JINR,
141980 Dubna, Russia\\
$^{2}$ National Research Nuclear University MEPhI (Moscow Engineering Physics Institute), \\
Kashirskoe highway 31, Moscow, 115409, Russia\\
}

\date{Accepted 2021 October 1. Received 2021 September 29; in original form 2021 August 27}

\pubyear{2021}

\begin{document}
\label{firstpage}
\pagerange{\pageref{firstpage}--\pageref{lastpage}}
\maketitle

\begin{abstract}
Two groups of astronomers used the large telescopes Keck and VLT for decades to observe trajectories of bright stars near the Galactic Centre.
Based on results of their observations the astronomers concluded that trajectories of the stars are roughly elliptical
and foci of the orbits are approximately coincide with the Galactic Centre position.
In  a last few years  a
 self-gravitating dark matter core--halo distribution was suggested by Ruffini, Arg{\"u}elles, Rueda (RAR) and  this  model was actively used in consequent studies. In particular, recently it  has been
  claimed that  the RAR-model provides a better fit
of trajectories of bright stars in comparison  to  the conventional model with a supermassive black hole.
 The  dark matter distribution with a dense core having a constant density as it was suggested in the RAR-model leaves trajectories of stars elliptical like in Kepler's two-body problem. However, in this case not the foci of the ellipses coincide with the Galactic Center but their centers while the orbital periods do not depend on semi-major axes.
 These properties are not consistent with the observational data for trajectories of bright stars.

\end{abstract}




\section{Introduction}

For decades astronomers observed bright stars which are moving very closely to the Galactic Centre. Analyzing trajectories of these stars one can deduce the functional form of the gravitational potential there. Observations of Keck and GRAVITY (VLT) groups showed that in the first approximation the stars are moving along elliptical orbits and foci of these orbits are roughly coinciding with the Galactic Centre \citep{2003ApJ...586L.127G,2005ApJ...620..744G,2008ApJ...689.1044G,2009ApJ...692.1075G,2010RvMP...82.3121G,2020Natur.577..337C}. If we apply the simplest
approach based on laws of classical mechanics we could follow the Newton's derivation of the gravitational potential from the Kepler's laws  \citep{Kopeikin_2011}. For the conventional model with the supermassive black hole and for a dense core model considered below such an approach looks reasonable since for these orbits radial coordinates $r >> r_g$ ($r_g$ is the  Schwarzschild radius for supermassive black hole) and general relativistic effects could be evaluated as perturbations of Newtonian model. Recently,    \citet{2018A&A...615L..15G,2019A&A...625L..10G} and Keck group \citep{2019Sci...365..664D} has reported discovery of the relativistic redshift of spectral lines for S2 star near its periapsis passage in May 2018, the redshift was consistent with theoretical estimates done
in the first post-Newtonian approximation of general relativity. These results are confirming the universality of gravity laws and did not support the alternative theories of gravity predicting different value of redshift in the weak and strong gravitational field \citep{2006PhRvD..73j4019C,2007MNRAS.375.1423C,2011PhR...509..167C}.
In the last year \citet{2020A&A...636L...5G} reported a discovery of relativistic (Schwarzschild) precession for S2 star orbit.

If there is a supermassive black hole along with a bulk distribution of ordinary matter forming a stellar cluster and a dark matter inside the orbits of bright stars, the bulk distribution of matter causes precession of the stellar orbits in the direction
which is opposite to that caused by relativistic effects of the black hole \citep{2001A&A...374...95R,2007PASP..119..349N,2007PhRvD..76f2001Z}.

Observational data for trajectories of bright stars can be used to test predictions of general relativity and  to constrain parameters of alternative theories of gravity such as $f(R)$ \citep{2012PhRvD..85l4004B},
Yukawa potential \citep{2013JCAP...11..050B}, theories with massive graviton \citep{2016JCAP...05..045Z,2018JCAP...04..050Z,2017PhRvL.118u1101H}, black holes with a tidal charge \citep{Zakharov_2018}, and others \citep{2019MNRAS.489.4606G}.
Constraints on variations of a fine structure constant were imposed by observations of bright stars near the supermassive black hole at the Galactic Centre as shown in \citep{2020PhRvL.124h1101H}, where other astrophysical ways to constrain the fine structure constant variations are mentioned as well.

Many theoretical models for the Galactic Centre have been proposed in the past and some of them were rejected as being inconsistent with the analysis of observational data and at the time being the model consisting of a single supermassive black hole looks the most preferable. Nonetheless, several years ago \cite{2015MNRAS.451..622R} proposed that the gravitational potential at the central part of our galaxy is  better approximated by a dark matter distribution having a dense core and a diluted halo.
 Later, the dark matter distribution was called the RAR-model. Recently \citet{2021MNRAS.505L..64B} declared that this model provides a better fit of trajectories of the bright stars in comparison with the supermassive black hole model. Below we consider properties of bright star trajectories in gravitational field of a dense core in the RAR-model and conclude that these properties are not consistent with the existing observational data.

\section{Central field}

We assume that a gravitational potential for the Galactic Centre is spherically symmetric. At this stage we use  the laws of the Newtonian mechanics.  This is sufficient for our goal and the general relativistic effects have a post-Newtonian order of magnitude and do not change our conclusions.
Here we introduce basic notations for the central field following  the textbooks  by \citet{Whittaker_1917,Arnold_1989, Goldstein_2002}. Let's assume, that the potential $U=U({r})$ does not depend on vector $\bf {r}$, it depends only on the radial distance $r=|{\bf r}|$. Then, motion of a freely-falling test particle with mass $m$ obeys the Newtonian law of gravity
\begin{equation}
  m \ddot{\bf{r}} =-\dfrac{\partial U}{\partial \mathbf{r}} = -\dfrac{\partial U}{\partial {r}}\mathbf{e}_r,
  	\label{CF_potential}
\end{equation}
where  $\mathbf{e}_r$ is an orthonormal vector in a radial direction.
Therefore,  the conservation law for angular momentum can be written in the form
\begin{equation}
h=m\dot{\phi} (t) r^2(t)=const,
	\label{CF_potential_0}
\end{equation}
where $\phi$ is the angle in the orbital plane and the dot denotes a time derivative.
Introducing potential $V(r)=U(r)+\dfrac{h^2}{2mr^2}$, one has
\begin{equation}
    m\ddot r =-\dfrac{\partial V}{\partial {r}}.
	\label{CF_potential2}
\end{equation}
The total energy of the particle in one-dimensional case is
\begin{equation}
    E =\dfrac{m\dot{r}^2}{2}+V(r).
	\label{CF_potential3}
\end{equation}
From Eq. (\ref{CF_potential3})  one has
\begin{equation}
    \dot{r}=\sqrt{\dfrac{2}{m}\left[E-V(r)\right]},
	\label{CF_potential6}
\end{equation}
therefore,
\begin{equation}
    t=\int\dfrac{dr}{\sqrt{\dfrac{2}{m}\left[E-V(r)\right]}}.
	\label{CF_potential7}
\end{equation}
Integrating Eq. (\ref{CF_potential_0}) we get
\begin{equation}
    \phi=\int_{r_0}^{r}\dfrac{dr}{r^2\sqrt{\dfrac{2mE}{h^2}-\dfrac{2mU}{h^2}-\dfrac{1}{r^2}}}.
	\label{CF_potential9}
\end{equation}
If we introduce a new variable $u=1/r$, we have
\begin{equation}
    \phi=-\int_{u_0}^{u}\dfrac{du}{\sqrt{\dfrac{2mE}{h^2}-\dfrac{2mU}{h^2}-u^2}}.
	\label{CF_potential10}
\end{equation}
From inequality $V(r) \leq E$, we have one or several intervals \citep{Arnold_1989}
\begin{equation}
    0 \leq r_{min} \leq r \leq r_{max} \leq \infty,
	\label{CF_potential11}
\end{equation}
where $V(r_{min})=V(r_{max})=E$.
If
\begin{equation}
    0 < r_{min} < r_{max} < \infty,
	\label{CF_potential11a}
\end{equation}
the corresponding motion is bounded and it is located inside a ring with an inner radius $r_{min}$ and an outer radius $r_{max}$.

For bounded orbits we could introduce the following angle  characterizing precessional motion of the orbit, see for instance, the textbooks by \cite{Landau_1993} and \cite{Arnold_1989}
\begin{equation}
    \Phi=\int_{r_{min}}^{{r_{max}}}\dfrac{hdr}{mr^2\sqrt{\dfrac{2}{m}\left[E-V(r)\right]}}.
	\label{CF_potential12}
\end{equation}

For a central potential $U(r)$ \citet{Bertrand_1873} proved the following theorem (see also the available English translation  by \citet{Santos_2011}):
there are only two potentials of the central field which yield closed elliptical orbits.
These potentials are the $U_{HO}=ar^2$ and $U_N=-k/r$ (the harmonic oscillator and the Newtonian potentials).
This result is also discussed by \cite{Landau_1993} and \cite{Arnold_1989}, while the book by \cite{Arnold_1990} presents a historical background leading to the discovery of Bertrand's theorem.

\section{Closed orbits in harmonic oscillator potential}
As soon as a particular potential is known we can calculate the angle
$\Phi$ in Eq. (\ref{CF_potential12}).
For instance, for the harmonic oscillator and for this case $\Phi_{HO}=\pi/2$ \citep{Arnold_1989} while it is well-known that for the Newtonian potential one has
$\Phi_{N}=\pi$. It is known that the orbit is closed if the angle $\Phi$ commensurable with $2 \pi$, or in  other words, if $\Phi=2\pi m/n$, where $m$ and $n$ are integers \citep{Arnold_1989}. Thus, computation of the angle $\Phi$ immediately tells us whether the orbit is closed or not. Clearly, they are closed in both cases of the harmonic oscillator and the Newtonian potential.
What remains is to understand where the  geometric center of the elliptical orbit is located with respect to the center of gravity. It is well known that for the Newtonian potential the center of gravity is not at the center of the ellipse but in its focal point. This is not the case for the harmonic oscillator potential as we demonstrate below.

Substituting $U_{HO}=a/u^2$ in Eq. (\ref{CF_potential10}), we obtain (since $u=1/r$)
\begin{equation}
    \phi=-\int_{u_0}^{u}\dfrac{du}{\sqrt{\dfrac{2mE}{h^2}-\dfrac{2ma}{u^2h^2}-u^2}}.
	\label{Orbits_1}
\end{equation}

Following \citet{Whittaker_1917} we introduce notations   $\beta=\dfrac{2mE}{h^2}$, $\mu= 2ma$, $v=u^2$ and showed that
   \begin{equation}
    \phi=-\dfrac{1}{2}\int_{v_0}^{v}\dfrac{dv}{\sqrt{\beta v-\dfrac{\mu}{h^2}-v^2}}.
	\label{Orbits_2}
\end{equation}

The integral in Eq. (\ref{Orbits_2}) is evaluated with Euler's substitutions
and if  $\gamma$ is an integration constant, we obtain   \citep{Whittaker_1917}
  \begin{equation}
    2(\phi-\gamma)=\arccos{\dfrac{v-\dfrac{\beta}{2}}{\sqrt{\dfrac{\beta^2}{4}-\dfrac{\mu}{h^2}}}},
	\label{Orbits_3}
\end{equation}
therefore,  following \citet{Whittaker_1917}
\begin{equation}
    \dfrac{1}{r^2}=\dfrac{\beta}{2}+\sqrt{\dfrac{\beta^2}{4}-\dfrac{\mu}{h^2}}\cos(2\phi-2\gamma),
	\label{Orbits_4}
\end{equation}

This equation  describes the ellipse whose center is at the origin of coordinates that coincides with the center of gravity.
The angular frequency and the period of harmonic oscillator potential are $\omega=\sqrt{2a}$ and   $T_{HO}= 2\pi/\omega$.

\section{Central part of dark matter distribution in the RAR model: core potential}

Recently,   many alternative models have been proposed for the Galactic Center in addition to the conventional model of the supermassive black hole
with mass around $4 \times 10^6 M_\odot$. There are two main reasons to develop such alternative models. First, some theorists hope that
alternative theories of gravity could explain dark matter phenomenon and properties of compact objects in alternative theories may be different from properties of black holes in general relativity. Second, other researchers hope that proposed dark matter distributions could provide better fits for observational data and many authors argued that instead of the supermassive black hole  the Galactic Centre contains  a dense core with constant density, see for instance,
\citet{1998MNRAS.296..569C,2002PrPNP..48..291B,Bilic_2002b,2015MNRAS.451..622R,2016JCAP...04..038A,2018PDU....21...82A,2019PDU....24..278A,2020MNRAS.tmp.3770A,2020A&A...641A..34B,2021MNRAS.505L..64B}
and references therein. These authors considered various dark matter distributions with a low density halo and a high density core. The most popular model of this type was proposed  by \citet{2015MNRAS.451..622R} and it is called the RAR-model. In the RAR model the central part of the galaxy represents a ball of constant density $\rho_0$ having a radius $R$.
Let us consider motion of a test particle (a star) in the potential field of the RAR model. We assume that the orbit of the particle is bounded and has a pericenter $r_p$ and an apocenter $r_a$.

In this case, one could analyze three different conditions for density and related potentials.

\underline{Condition A}.  $r_a \leq R$ and star is moving inside the ball of the constant density  $\rho_0$.
Then we have a harmonic oscillator potential  $U_{HO}(r)$ along the entire orbit, since according to the law of the Newtonian gravity the potential is formed only by the matter contained inside the radius $r$
\begin{eqnarray}
   4\pi G  \int_{0}^r\rho_0 r dr = U_{HO}(r)=ar^2, 
	\label{Density_0}
\end{eqnarray}
where $a=2 \pi G  \rho_0$, $G$ is the gravitational constant.

\underline{Condition B}.  $r_p \leq R < r_a$ and the particle moves both inside and outside of the dense core.

\underline{Condition B1}. If we assume that the mass of the dark matter contained inside of the pericenter of the particle's orbit is equal to the mass of the presumed BH
\begin{eqnarray}
   4\pi \int_0^{r_p}\rho_0 r^2 dr \approx M_{BH}
	\label{Density_2}
\end{eqnarray}
and the mass of the dark matter outside of the ball of radius $r_p$ is much smaller than the mass of the BH
\begin{eqnarray}
   4\pi \int_{r_p}^r\rho(r)r^2 dr << M_{BH},
	\label{Density_3}
\end{eqnarray}
then this case is very similar to the standard approach consisting of a supermassive black hole and perhaps an extended matter distribution with a negligibly small mass inside a quasi-elliptical orbit of a star.
As we noted earlier for our needs the Newtonian approximation looks reasonable since currently the observed bright stars move very far away from the black hole horizon.

\underline{Condition B2}. If
\begin{eqnarray}
   4\pi G  \int_{0}^r\rho(r)r dr =\tilde{U}_{HO}(r) \approx U_{HO}(r)=ar^2, \quad \forall r \in (R, r_a),
	\label{Density_4}
\end{eqnarray}
 then one could use $U_{HO}(r)$ as a suitable approximation for the gravitational potential. Eq. (\ref{Density_4}) is valid for very small density gradients for all $r$ in the interval $R < r < r_a$ (or in other words, for these $r$ we have $\rho(r) \approx \rho_0$).  This case is similar to the case with Condition A, where $U_{HO}(r)$ determines the exact expression for the gravitational potential, however, trajectories in the harmonic oscillator potential are not exact solutions but suitable approximations for potential  $\tilde{U}_{HO}(r)$.

\underline{Condition  B3}. If conditions B1 and B2 are not valid, then the theoretical model for gravitational potential can not be approximated
  with   $U_{HO}(r)$ and $U_{N}(r)$. Therefore, in principle there is an opportunity that there are bounded orbits which are not elliptical and not closed.
  One should note that  observed bounded orbits at the Galactic Center are closed and quasi-elliptical in the first approximation. This case deserves a separate detailed investigation but we do not pursue it in this Letter.

  \underline{Condition  C}. In this case $ R < r_p$ and
relations (\ref{Density_2},\ref{Density_3}) are valid   then
$U_{N}(r)$ could be chosen as a good approximation for a gravitational potential. Otherwise, similarly to case B3, gravitational potential can not be approximated   with   $U_{HO}(r)$ and $U_{N}(r)$.

\section{Conclusions}
High-precision astrometric monitoring of quasi-elliptical trajectories of stars with high eccentricities is instrumental for distinguishing  $U_{HO}(r)$ and $U_{N}(r)$
potentials since  the centers of the elliptic orbits of the stars
   should coincide with the Galactic Center in the case of the RAR potentials for a dense core while in the case of the Newtonian potential star foci of the ellipses coincide with the Center.
As it was noted in Section 3 the orbital periods of stars moving in  the field of harmonic oscillator potential are constant and they do not depend on semi-major axes.
Assume that we have to choose a suitable potential for the Galactic Centre from two options $U_{HO}(r)$ and $U_{N}(r)$. An inspection of a set of trajectories with high eccentricities (see, for instance, Fig. 16 in paper by \cite{2009ApJ...692.1075G}, Fig. 16 (right panel) from paper by
  \cite{2010RvMP...82.3121G} and Fig.8 in paper by \cite{2017ApJ...837...30G})
clearly showed that stars are moving around the joint focus but not the centre.

 Therefore, currently,  we can state with a rather good confidence that all bounded observed orbits of bright stars near the Galactic Centre are elliptical and their foci  practically coincide with the Galactic Centre.  Therefore, we have to conclude that  the central potential
 inside the spherical shell where these observed bounded orbits are located must be the Newtonian
 $U_{N}(r)$. In Section 3 it was shown that in the case of a harmonic oscillator potential $U_{HO}(r)$ which corresponds to the RAR dark matter  model with a dense core inside a ball with a constant density and ellipse centers for different orbits should coincide with the Galactic Centre. Periods of trajectories in $U_{HO}(r)$ potential do not depend on semi-major axis. These properties are not supported by observational data reported by Keck and GRAVITY (VLT) teams which means that the conventional model with a supermassive black hole  generating approximately the Newtonian potential
 in a spherical shell where bright star orbits are located, is a preferable model for physical interpretation of the observed trajectories.

\section*{Acknowledgements}

The author thanks G. A. Alekseev for useful discussions. The author also appreciates a referee for numerous useful remarks which helped to improve the Letter.

\section*{Data Availability}

The astrometric data used in this letter were obtained from papers by \citet{2009ApJ...692.1075G,2017ApJ...837...30G,2010RvMP...82.3121G,2019Sci...365..664D}.






%
%


\bsp	
\label{lastpage}
\end{document}